\documentclass[aps,pre,twocolumn, amsmath,amssymb,showpacs,superscriptaddress,floatfix]{revtex4-1}
\usepackage{graphicx}

\usepackage{color}

\begin{document}
\title{Polymer-mediated entropic forces between scale-free objects}

\author{Mohammad F. Maghrebi}
\email{Email: magrebi@mit.edu}
\affiliation{Massachusetts Institute of Technology, Department of
  Physics, Cambridge, Massachusetts 02139, USA}
\author{Yacov Kantor}
\affiliation{Raymond and Beverly Sackler School of Physics and
Astronomy, Tel Aviv University, Tel Aviv 69978, Israel}
\author{Mehran Kardar}
\affiliation{Massachusetts Institute of Technology, Department of
  Physics, Cambridge, Massachusetts 02139, USA}

\date{\today}

\pacs{64.60.F- 
82.35.Lr 
05.40.Fb 
}

\begin{abstract}
The number of configurations of a polymer is reduced in the presence of
a barrier or an obstacle. The resulting loss of entropy adds a repulsive
component to other forces generated by interaction potentials.
When the obstructions are scale invariant shapes (such as cones, wedges,
lines or planes) the only relevant length scales are the polymer size $R_0$
and characteristic separations, severely constraining the functional form
of entropic forces. Specifically, we consider a polymer (single strand or
star) attached to the tip of a cone, at a separation $h$ from a surface
(or another cone). At close proximity, such that $h\ll R_0$, separation
is the only remaining relevant scale and the entropic force must take the
form $F={\cal A} \, k_{B}T/h$. The amplitude $\cal A$ is universal, and
can be related to exponents $\eta$ governing the anomalous scaling of
polymer correlations in the presence of obstacles. We use analytical,
numerical and $\epsilon$-expansion techniques to compute the exponent
$\eta$ for a polymer attached to the tip of the cone (with or without
an additional plate or cone) for ideal and self-avoiding polymers. The
entropic force is of the order of 0.1pN at 0.1$\mu$m for a single polymer,
and can be increased for a star polymer.

\end{abstract}
\maketitle

\section{Introduction}
A host of single molecule manipulation
techniques~\cite{bustamante03,kellermayer,neuman,deniz,neuman08}, employing
atomic force microscopes (AFMs)~\cite{fisher}, microneedles~\cite{kishino},
optical~\cite{neuman2004,hormeno} and magnetic~\cite{gosse} tweezers, have
provided quite detailed studies of shapes and forces in long polymers.
The positional accuracy of an AFM tip~\cite{kikuchi97,neuman08} can be as good
as a few nm, while  forces of order of 1~pN can be measured, with measurements
carried out in nearly biological conditions~\cite{brk1997,drake89}.
The main thrust of current experimental research is to extract specific
information about molecular shapes and reactions from force-displacement curves.
The interpretation of such data is complicated by a host of factors such as
interactions of the molecule with the probes, and non-equilibrium effects arising
from rapid and large imposed deformations.
Nonetheless, much recent activity is devoted to obtaining the equilibrium
free energy landscape of long molecules by processing results from repeated
non-equilibrium measurements  (see, e.g., Refs. \cite{Bustamante05,Maragakis08,Peri11}).
Here, we consider another contribution to the force arising from the loss of entropy
in the presence of constraints imposed by the probes.
While most likely weaker than forces due to interaction potentials, the enhanced
AFM  sensitivities are indeed approaching the range where entropic forces of
long polymers in a solvent can be significant  (even when the deformation of
the polymer is relatively slight).
From the theoretical perspective, the attraction of entropic forces on polymers is
their independence of microscopic details~\cite{degennesSC}.

At a finite temperature $T$, a long homo-polymer in a good solvent
fluctuates between a large number of configurations, and its `macroscopic' properties
(such as size and elasticity beyond a characteristic persistence length) are
governed by entropic considerations.
To illustrate entropic elasticity, consider the hypothetical situation of a
polymer with one end fixed to the origin, while the other end
explores allowed positions at $\vec{r}$. In equilibrium, due to rotational
symmetry $\langle\vec{r}\,\rangle=0$. The symmetry is broken if a weak force
$\vec{F}$ is applied by a point-like probe at the end of the polymer. Then,
in $d$-dimensional space $\langle\vec{r}\,\rangle=(R_0^2/dk_BT)\vec{F}$, where
$R_0^2$ is the mean squared end-to-end distance in the absence of external
force~\cite{degennesSC}. This is mostly a realignment, without substantial
change of polymer conformations for $F\ll F_{\rm cr}\equiv dk_BT/R_0$.
For $R_0$=0.1$\mu$m at room temperature $F_{\rm cr}\approx$ 0.1pN is the crossover
value that separates weak and strong forces; stronger force ($F\gg F_{\rm cr}$) leads
to substantial stretching of the polymer and associated reduction in entropy.
Similar considerations come into play when the attachment points are replaced by
macroscopic obstacles (probes): Under a strong {\it stretching} force polymer
configurations are not very much influenced by the shapes of the probes, and
convenient models (such as Ref.~\cite{Marko95}) have been developed for the
resulting force-extension relations. Here, we focus on the opposite limit of strong
{\it compression} in which the loss of entropy, and the corresponding force, are
dominated by confinement within boundaries.

In this work we consider setups where a polymer is attached by one end to an object
(the probe), while the other end is free.
The probe is then brought to the vicinity of a second solid obstacle, and the loss of polymer entropy
leads to a force between the two objects.
In particular, we consider confining (probe and obstacle) shapes that are
self-similar (at least on scales comparable to $R_0$) at separations
$h\ll R_0$.
Possible setups include sharp AFM tips approaching
flat surfaces covered with polymers, as in experimental
studies~\cite{Oshea93,Lea94,Overney96,Kelley98,Mendez09} of polymer brushes.
Entropic considerations in a dense brush must account for interactions between polymers~ \cite{Netz03},
complicating the calculation of forces exerted on the AFM tip~\cite{Halperin10}.
Here we consider the simpler cases of a single linear or star polymer, avoiding the dense limit.
For $h\ll R_0$ dimensional considerations
suggest the particularly simple force-separation relation $F={\cal A} \, k_{B}T/h$,
where the `universal' amplitude ${\cal A}$ only depends on basic geometrical
properties, and gross features of the polymer.

In a previous paper we outlined some results pertaining to entropic interactions,
and critical exponents, for a polymer attached to a cone~\cite{MKK_EPL96}.
In this paper we expand on the details of the calculations, as well as
generalize and expand the results to other setups. Section~\ref{sec:forces}
demonstrates how the force amplitude ${\cal A}$ is related to the exponent $\eta$,
characterizing the anomalous scaling of polymer correlations. The latter exponent
depends on the shape of confining boundaries, and is the focus of computations
that follow. In Sec.~\ref{sec:scaleinvariant} we outline the general approach
to computing $\eta$ for {\it ideal polymers} near scale-invariant surfaces,
while in Sec.~\ref{sec:idealcones} we make explicit computations for cones and
plates. The case of {\it self-avoiding} linear polymers in $d=3$ is  treated
by Monte Carlo simulations in Sec.~\ref{sec:SAW}. Perturbative $\epsilon=4-d$
expansions are employed in Sec.~\ref{sec:epsilon} to compute $\eta$ for
linear and star self-avoiding polymers attached to cone tips, or to a contact
point between a cone and a plate.

\section{Polymer-mediated forces}\label{sec:forces}
Let us first consider an idealized setup in which a polymer is attached to
the tip of a solid cone, approaching a solid plate (or another cone).
This exemplifies a geometry of obstacles in which the only (non-microscopic)
length scale is provided by the tip-plate (or tip-tip) separation $h$.
The polymer itself undergoes self-similar fluctuations, spanning length
scales  intermediate between microscopic (monomer size, or persistence
length $a$) and macroscopic. The latter is set by the typical end-to-end
distance $R_0$, or by the radius of gyration $R_g$ (which differs from
$R_0$ only by a multiplicative factor of order of unity). The typical size
of the polymer grows with the number of monomers $N$ through the scaling
relation $R_0\sim R_g\propto N^\nu$. The self-similar shape of the obstacles
also presumably extends only up to a characteristic scale $H$, say the
height beyond which the cone is terminated or changes shape.

Neglecting any structure (and hence energy) associated with the polymer, variations in
free energy as the cone (plus tip-attached polymer) approaches the plate are entirely entropic
in origin and proportional to $k_BT$.
The entropic force has dimensions of energy divided by length, and at separations
$a\ll h\ll R_0\ll H$  the relevant length scale is $h$, and it is reasonable to posit
\begin{equation}
\label{Eq:force}
F={\cal A}\,\frac{k_BT}{h}\ .
\end{equation}
This is because the polymer configurations are constrained only on the scale of confinement $h$; increasing the length of the polymer or the size of the cone
(as long as $h\ll R_0\ll H$) should not influence the force. Furthermore the entropy change is independent of $a$.

Indeed, the simple force law of Eq.~(\ref{Eq:force}) applies to all circumstances where the separation provides the only relevant length scale, and   follows easily from various polymer
scaling forms (see, e.g. the derivation below) such as in
Refs.~\cite{Eisenriegler82,Duplantier86,Row11}.
The dimensionless amplitude $\cal A$ will depend on geometric factors characterizing
the confining boundaries such as the opening angle of a cone $\Theta$ (and if tilted, on the
corresponding angle).
It will also depend on factors that characterize the scaling of polymeric fluctuations,
thus differing in cases of ideal and self-avoiding polymers, and for linear, star, and brush polymers.
In the following we shall demonstrate that in a number of setups, the amplitude $\cal A$
can be related to variations of the (universal, but shape dependent) exponent $\eta$, characterizing polymeric correlations.

\begin{figure}
\includegraphics[width=8cm]{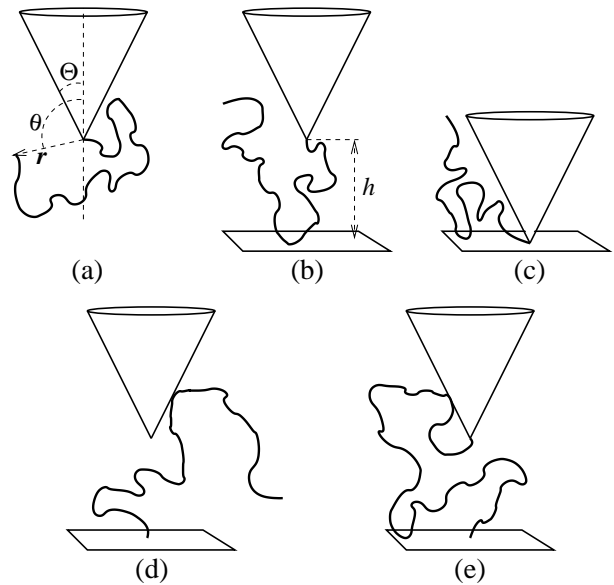}
\caption{(a) Polymer attached to the tip of a solid  cone with apex
semi-angle $\Theta$ (configuration ``c"); positions are described by the spherical
coordinates $r,\theta$ and azimuthal angle $\phi$ (not shown).
(b) The tip, where the polymer is attached, is at a distance $h\ll R_0$
from the plate.
(c) The tip touching the plate (configuration ``cp").
(d) Tip is at a finite distance from a plate to which the polymer is attached.
(e) Polymer attached to both surfaces.
\label{fig:def_geometry}}
\end{figure}

A chain starting at the tip of a cone provides a first approximation to a polymer linked
to the tip of an AFM probe as depicted in Fig.~\ref{fig:def_geometry}a.
With the cone far away from a plate ($h\gg R_0$), the number of configurations of
the polymer grows with the number of monomers as
\begin{equation}\label{Eq:gammadef}
{\cal N}_{\rm c}= b \, z^N  N^{\gamma_{\rm c}(\Theta)-1},
\end{equation}
where the effective coordination number $z$, as well as the pre-factor $b$,
depend on the microscopic details (such as the scale $a$),
while the  `universal' exponent $\gamma_{\rm c}$  only
depends on the cone angle.
When the cone touches the plate as in Fig.~\ref{fig:def_geometry}c, the number of
configurations is reduced to ${\cal N}_{\rm cp}$ with the same form as Eq.~\eqref{Eq:gammadef},
but with a different exponent $\gamma_{\rm cp}(\Theta)$.
We shall henceforth use an exponent subscript `${\rm s}$' (as in $\gamma_{\rm s}$)
to refer to the above cases, with ``s=c" for cone and ``s=cp" for cone+plate;
the absence of a subscript (as in $\gamma$) will signify a free polymer.
The work done against the entropic force in bringing in the tip from
afar to contact the plate can now be computed  from Eq.~\eqref{Eq:force} as
\begin{equation}
\label{Eq:work}
W=\!\int_a^{R_0}\!\!{\rm d}h~{\cal A}\frac{k_BT}{h}\ =\!{\cal A}k_BT\ln\frac{R_0}{a}
 ={\cal A}\,\nu k_BT\ln N.
\end{equation}
This work can also be computed from the change in free energies between the final
and initial states, due to the
change in entropy, as
\begin{equation}
\label{Eq:DF}
\Delta {\cal F}= -T\Delta{\cal S}=T{\cal S}_{\rm c}-T{\cal S}_{\rm cp}
=k_BT(\gamma_{\rm c}-\gamma_{\rm cp})\ln N  \,,
\end{equation}
with the entropy ${\cal S}=-k_B\ln{\cal N}$  computed from Eq.~(\ref{Eq:gammadef}).
By equating $W$ and $\Delta {\cal F}$ we find
\begin{equation}
\label{Eq:A}
{\cal A}=\frac{\gamma_{\rm c}-\gamma_{\rm cp}}{\nu}=\eta_{\rm cp}-\eta_{\rm c}\, ;
\end{equation}
the final result obtained from the scaling law~\cite{degennesSC}
\begin{equation}
\label{Eq:gammanu}
\gamma_{\rm s}=(2-\eta_{\rm s})\nu,
\end{equation}
where the exponent $\eta$ characterizes the anomalous decay of correlations ($\sim 1/r^{d-2+\eta}$).

\begin{figure}
\includegraphics[width=8cm]{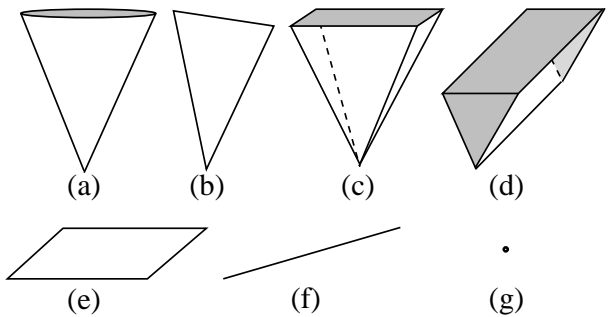}
\caption{
Examples of three dimensional figures without a length scale;
grey surfaces indicate truncation only for graphical representations.
(a) circular cone,
(b) two-dimensional sector of a circle (in three-dimensional space), (c) pyramid,
(d) wedge, (e) plane, (f) line, (g) point.
A polymer is to be attached to the point (g), to the apex of (a), (b), (c);
or to any point on the edge of (d), or the entirety of (e) or (f).
The plane and line can also be semi-infinite
with a polymer attached to their edge or end-point.
\label{fig:figselection}}
\end{figure}

The arguments presented in the previous paragraph rely only on the fact that
 configurations of the obstacles lack a length scale for both $h\to0$ and $h\to\infty$.
Consequently, similar reasoning can be applied to a variety of objects such as those depicted
in Fig.~\ref{fig:figselection}, or  combinations of such objects. The force prefactor is then related
to the exponents in the initial and the final states by
\begin{equation}
\label{Eq:Agen}
{\cal A}=\frac{\gamma^{\rm initial}-\gamma^{\rm final}}{\nu}
=\eta^{\rm final}-\eta^{\rm initial}\, .
\end{equation}
Consider, for instance, an ideal polymer in free space (or held by a
point-like object (Fig.~\ref{fig:figselection}g)). Its number of configurations
is $z^N$, i.e. $\gamma\equiv\gamma_0=1$, and correspondingly $\eta_0=0$,
since $\nu=\nu_0=1/2$. Subscript 0 will henceforth indicate exponents describing
ideal polymers.
If the end of the polymer is brought into contact with a plane, then by the method
of images \cite{chandra} it can be shown that in
any space dimension $d$ the number of configurations scales as $z^N N^{-1/2}$,
i.e. with $\gamma\equiv\gamma_{{\rm p}0}=1/2$, and  corresponding to
$\eta_{{\rm p}0}=1$. Thus for a long polymer held by a point-like object
at a distance $h$ from a plane, the entropic force in Eq.~\eqref{Eq:force}
will have a prefactor ${\cal A}=1$. This result is valid for any $d$.
For self-avoiding polymers in $d=2$, the exponents are $\gamma=43/32$,
$\gamma_{\rm p}=61/64$~\cite{cardy1} and $\nu=3/4$~\cite{Cardyconf} leading
to ${\cal A}=25/48\approx0.52$, while in $d=3$, we get
$\gamma\approx1.158$~\cite{caracciolo}, $\gamma_{\rm p}\approx0.70$ and
$\nu\approx0.59$~\cite{debell} leading to ${\cal A}\approx0.78$.
Similar exact results can be derived for more complex self-avoiding
polymers, such as star polymers or branched polymers of arbitrary
topology in $d=2$~\cite{Duplantier86}.

Note that the force amplitude does depend on the surface to which the polymer
is attached. For example, consider the configuration in Fig.~\ref{fig:def_geometry}d
where the polymer is attached to the plate. The initial configuration
corresponds to a polymer attached to an infinite surface, thus
$\eta^{\rm initial}=\eta_{\rm p}$, while the final configuration is
the same as in Fig.~\ref{fig:def_geometry}c, with $\eta^{\rm final}=\eta_{\rm cp}$.
This leads to ${\cal A}=\eta_{\rm cp}-\eta_{\rm p}$, as opposed to
${\cal A}=\eta_{\rm cp}-\eta_{\rm c}$ for the case that the polymer is
attached to the cone. There are a number of detailed studies of
small surfaces (such as AFM tips)~\cite{Murat96,Jimenez98,Subra95,Guffond97,Sevick99,Sevick00,Hsu07}
compressing a single polymer (or few polymers) grafted to a surface.
However, all these studies concern {\em non}-scale invariant geometries.

When the obstacles are separated by $h$, the loss of polymer entropy (and the
corresponding pressure leading to the force in Eq.~\eqref{Eq:force}) is
concentrated in the confinement region. The part of the polymer that wanders
away from this area is relatively unperturbed and does not contribute to
the force. Therefore for large $N$, Eq.~\eqref{Eq:force} is independent of
polymer size. This is not the case when the entire polymer contributes to
the force,  as for a polymer held between two plates, where the  force is
proportional to $N$. Thus, the argument fails when dimensionality of the
system is changed between the initial and final states. For example when a
3-dimensional polymer connected to a two-dimensional plane approaches a
parallel plane, in the final configuration the confined polymer is
essentially two-dimensional. Free energies in the initial and final state
have different extensive parts, and the polymer-mediated force depends on
the number of monomers \cite{Schlesener01}.

It is worth reiterating the idealizations leading to Eqs.~\eqref{Eq:force}
and~\eqref{Eq:Agen}.
Real obstacles are self-similar over a range of length scales; for example an
AFM tip may be rounded, or abruptly change its angle.
Equation~\eqref{Eq:force} will then be applicable to a corresponding
interval, and there may be  non-trivial crossovers between the different
regimes~\cite{bubis}.
By focusing on entropy, we have assumed that the only
interaction between the polymer and the obstacles is due to hard-core
exclusion. Attractive interactions between the polymer and surface will
introduce temperature dependent corrections, and additional size
scales~\cite{degennesSC}.
Weak interactions are asymptotically irrelevant, but strong interactions
may lead to a phase in which the polymer is absorbed onto the obstacles~\cite{Eisenriegler82},
rendering the entropic considerations presented here inappropriate.
At high enough temperatures we expect entropic interactions to be
dominant \cite{degennesSC}.
However, the absorption of AFM constrained polymers to attractive probes
is indeed an interesting topic which falls outside the purview of this paper.

\section{Ideal polymers near scale-invariant surfaces}\label{sec:scaleinvariant}

Having reduced the computation of the force  to calculation of correlation functions,
the remainder of the paper is dedicated to computing exponents characterizing polymer correlations.
The behavior of idealized
polymers, i.e. in the absence of self-avoidance and other interactions,  is closely related to the diffusion problem: the spatial configuration of a polymer of $N$-segments
can be regarded as the space--time trajectory of a random walker viewed at
integer times up to $t=N$.
Summing over all trajectories starting at  ${\bf r}$ and ending at ${\bf r}'$ is then equivalent
to considering the probability of diffusion from ${\bf r}$ to ${\bf r}'$ in time $t$, which
satisfies the diffusion equation
\begin{equation}
\label{Eq:diffusionP}
\frac{\partial P({\bf r},{\bf r}',t)}{\partial t}=D{\nabla'}^2 P({\bf r},{\bf r}',t).
\end{equation}
The prime sign on the Laplacian indicates the spatial derivatives
with respect to components of ${\bf r}'$, and the initial condition
is $P({\bf r},{\bf r}',t=0)=\delta^d({\bf r}-{\bf r}')$.
The diffusion constant $D$ is chosen such that in free space the mean squared distance
coincides with the random walk value of $\langle({\bf r}-{\bf r}')^2\rangle=a^2N=2dDt$ in $d$ space dimensions, and thus $D=a^2/2d$.
If a polymer is confined by {\em repulsive} walls, its configurations
correspond to paths of a random walker that does no cross
these boundaries. In the language of diffusion, the exclusion of
paths that cross the boundaries is accomplished by assuming that they
are {\em absorbing} surfaces.

To compute variations in the number of states, we are interested in the total
survival probability
$S({\bf r},t)\equiv\int P({\bf r},{\bf r}',t){\rm d}^d{\bf r}'$.
which also satisfies the equation~\cite{weiss_book}
\begin{equation}
\label{Eq:diffusion}
\frac{\partial S({\bf r},t)}{\partial t}=D\nabla^2 S({\bf r},t),
\end{equation}
with absorbing boundary conditions.  The initial condition
for survival probability is $S({\bf r},t=0)=1$, everywhere inside the
space where the particle can diffuse, and $S({\bf r},t)=0$ on the
absorbing boundaries.
(While the space and time variables are continuous in the above equation,
appropriately discretized equations can be applied to diffusion or random walk on a lattice.)

In the absence of confinement, the number of configurations ${\cal N}$ of an
ideal polymer grows exponentially with its length $N$. On a regular discrete
lattice, ${\cal N}$ is simply related to lattice coordination number $z$ by
${\cal N}=z^N$. The presence of repulsive boundaries reduces the number of
configurations to
\begin{equation}
\label{Eq:NvsS}
{\cal N}({\bf r},N)=z^N S({\bf r},kN),
\end{equation}
where the prefactor $k=a^2/(2dD)$ converts  the dimensionless number
of steps $N$ to time $t$ in the diffusion equation.
Numerical solutions of Eqs.~(\ref{Eq:diffusionP},\ref{Eq:diffusion}) can be easily found for many
geometries, and analytical solutions are available for a number of simple shapes~\cite{Carslaw59}.
The generally complicated solutions simplify to some extent for the self-similar shapes
depicted in  Fig.~\ref{fig:figselection}.
In a coordinate system centered on a special point of the object
(such as tip of a cone) the dimensionless function $S$ can only depend on the
dimensionless vector ${\bf w}={\bf r}/\sqrt{Dt}$. Thus, $S({\bf r},t)=H({\bf w})$,
and Eq.~\eqref{Eq:diffusion} reduces to
\begin{equation}
\label{Eq:diffusion_red}
\nabla_{\mathbf w}^2 H+\frac{1}{2}{\bf w}\cdot\vec{\nabla}_{\mathbf w}H=0,
\end{equation}
where the subscript ${\bf w}$ indicates derivatives with respect to components of
${\bf w}$. In terms of these dimensionless variables, the function $H$ vanishes on the absorbing surfaces and increases to unity away from the surface (at separations of order one).
For some geometries, the solution to Eq.~\eqref{Eq:diffusion_red} can be
presented in terms of a radial distance $w$, and a combination of angular variables.
For example, generalized cones in $d$-dimensional space are described by a single polar angle
$\theta$ and $d-2$ azimuthal angles $\phi, \psi, \cdots$.
(While $d$-dimensional wedges can be treated similarly, it is more convenient to note that
the solution is independent of the coordinates parallel to the wedge, while
the cross-section perpendicular to the wedge resembles a cone.)
Furthermore, for $w\ll 1$ the distance dependence is expected to be a
simple power law $w^{\eta_0}\Psi(\theta,\phi,\dots)$. In this limit, the
second term in Eq.~\eqref{Eq:diffusion_red} becomes negligible, and the
problem reduces to solving the Laplace equation $\nabla_{\mathbf w}^2 (w^{\eta_0}\Psi)=0$.
The boundary conditions are now implemented solely in  terms of
angular variables, and the problem is reduced to finding a non-negative
function $\Psi$ that satisfies these conditions, and the corresponding $\eta_0$
for which such solution exists. The dependence of the latter value on the boundary shape
will be emphasized by denoting it as $\eta_{{\rm s}0}$. Since,
$S\sim w^{\eta_{{\rm s}0}}=(r/\sqrt{Dt})^{\eta_{{\rm s}0}}$ for a random walk
that starts a short distance $|{\bf r}|=a$ away from the cone, the survival
probability is $S\sim (a/\sqrt{Dt})^{\eta_{{\rm s}0}}$, and consequently the
number of configurations of an ideal polymer grows as
$z^N N^{-\eta_{{\rm s}0}/2}\equiv z^N N^{\gamma_{{\rm s}0}-1}$.
We see that $\gamma_{{\rm s}0}=1-\eta_{{\rm s}0}/2$ as in Eq.~\eqref{Eq:gammanu}
for $\nu_0=1/2$.

\section{Ideal polymers near cones and plates}\label{sec:idealcones}
As an example, consider a polymer attached to the apex of a cone, as in
Fig.~\ref{fig:def_geometry}a, or to a point of contact between a cone and
a plane, as in Fig.~\ref{fig:def_geometry}c.
When generalized to $d$-dimensions, this system has cylindrical symmetry,
and $\Psi$ is independent of the $d-2$ azimuthal angles.
Therefore, the Laplace equation simplifies to
\begin{equation}
\label{Eq:psi}
\frac{1}{(\sin\theta)^{d-2}}\frac{d}{d \theta}\left[(\sin\theta)^{d-2}
\frac{d\Psi}{d  \theta}\right]+\eta{_0}(d-2+\eta{_0})\Psi(\theta)=0\,,
\end{equation}
with an appropriate boundary condition on $\Psi(\theta)$. For an isolated cone,
the function $\Psi$ must be positive and regular outside the cone,
with  $d\Psi/d\theta|_{\theta=\pi}=0$ to avoid a cusp on the
symmetry axis, and $\Psi(\Theta)=0$ on the cone surface.
For the cone+plate, the appropriate solution is positive  and vanishes
at both $\theta=\Theta$ and $\theta=\pi/2$. The first case was considered
by Ben-Naim and Krapivsky~\cite{BenNaim} in connection with diffusion near
an absorbing boundary~\cite{redner_book}, and we follow their derivations.
The solution in general $d$ requires the use of associated Legendre
functions, but simplifies in a few cases described below.

\begin{figure}
\null\vskip 1cm
\includegraphics[width=7.5cm]{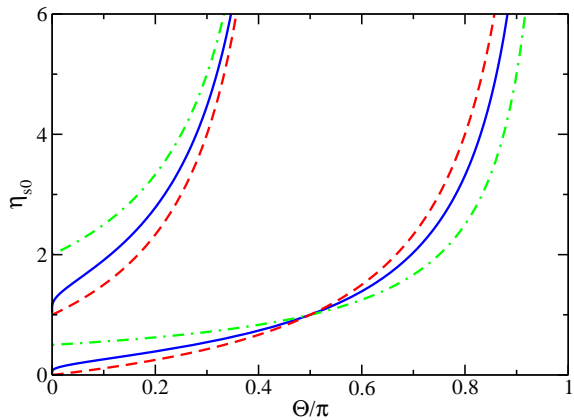}
\caption{
\label{fig:ThAll}
(Color online) The exponent $\eta_{\rm s0}$ for ideal polymers in $d=$2 (dot-dashed),
3 (solid), 4 (dashed) for cone (``s=c") of angle $\Theta$ (bottom curves), and
``s=cp" (top curves). (From Ref.~\protect{\cite{MKK_EPL96}}.)
}
\end{figure}

\par\noindent$\bullet$
{\it For $d=2$,} the problem of a cone coincides with that
of a wedge. In higher space dimensions the  wedge remains equivalent to
the two-dimensional case, as the function
$H$ is independent of the $d-2$ coordinates parallel to the wedge.
Thus, the following results for $d=2$ are also applicable to wedges
in any $d$. Equation~\eqref{Eq:psi} now reduces to
$\Psi''+\eta{_0}^2\Psi=0$; which is solved by  linear combinations of
$\sin(\eta{_0}\theta)$ and $\cos(\eta{_0}\theta)$.
The requirement that $\Psi$ is positive and regular, and vanishes on the object(s), yields
\begin{equation}
\label{Eq:beta2d}
\eta_{\rm c{0}}=\frac{\pi}{2(\pi-\Theta)},\quad{\rm and}\quad
\eta_{\rm cp{0}}=\frac{2\pi}{\pi-2\Theta}\,.
\end{equation}
Both results (depicted in Fig.~\ref{fig:ThAll}) go to a finite value as
$\Theta\to0$, reflecting the strong reduction in configurations due
to the remnant (barrier) line, and $\eta{_0}\to\infty$ when
the boundaries confine the polymer to a vanishing sector.

This computation can be easily generalized to the case of two cones (wedges)
(``s=cc") with apex semi-angles $\Theta_1$ and $\Theta_2$ touching
at their tips with a common symmetry axis. If one of the wedges
is tilted by an angle $\tau$ relative to that axis, then the survival
probability $S$ will be determined by the larger sector of free space.
The resulting exponent is
\begin{equation}
\label{Eq:eta_tilted}
\eta_{{\rm cc}0}=\frac{\pi}{\pi-(\Theta_1+\Theta_2-\tau)}.
\end{equation}
By proper choices of $\Theta$s and $\tau$ we can reproduce the results
of Eq.~\eqref{Eq:beta2d} as particular cases.

\par\noindent$\bullet$
{\it For $d=4$,} the substitution $\Psi=u/\sin\theta$ simplifies
Eq.~\eqref{Eq:psi} to $u''+(\eta{_0}+1)^2u=0$, solved by a linear
combination of $\sin[(\eta{_0}+1)\theta]$ and
$\cos[(\eta{_0}+1)\theta]$,
and we find
\begin{equation}
\label{Eq:beta4d}
\eta_{\rm c 0}=\frac{\Theta}{\pi-\Theta},\quad{\rm and}\quad
\eta_{\rm cp 0}=\frac{\pi+2\Theta}{\pi-2\Theta}\, ;
\end{equation}
depicted by the bottom and top dashed lines in Fig.~\ref{fig:ThAll}.
The cone exponent $\eta_{ {\rm c}0}$ vanishes linearly with $\Theta$
-- a needle in four dimensions is ``invisible" to a random walker.

As in the two-dimensional case, we may generalize to two touching cones
aligned along common axis with apex angles $\Theta_1$ and $\Theta_2$, with
\begin{equation}
 \label{Eq:eta4d_two}
\eta_{{\rm cc}0}=\frac{\Theta_1+\Theta_2}{\pi-(\Theta_1+\Theta_2)}.
\end{equation}
Unfortunately, when one of the cones is tilted, the problem loses its
azimuthal symmetry and Eq.~\eqref{Eq:psi} with its simple solutions
is no longer valid.

\par\noindent$\bullet$
{\it For $d=3$,} Eq.~\eqref{Eq:psi} becomes
\begin{equation}
\label{Eq:Legendre3D}
(1-\mu^2)\frac{d^2\Psi}{d\mu^2}-2\mu\,\frac{d\Psi}{d\mu}
+\eta_0(\eta_0+1)\Psi=0,
\end{equation}
where $\mu=\cos\theta$.
The general solution to this equation is given by regular (rather than associated)
Legendre functions
\begin{equation}
\label{Eq:3D_solution}
\Psi(\theta)=a_1P_{\eta_0}(\mu)+a_2Q_{\eta_0}(\mu).
\end{equation}
Note that $P_\alpha(1)=1$ for any $\alpha$, while $P_\alpha(-1)$ diverges for noninteger $\alpha$.
Similarly, $Q_\alpha(\pm1)$ is divergent. The linear combination in Eq.~\eqref{Eq:3D_solution}
can be made regular at -1 by a proper choice of $a_1/a_2$.
For the geometry described in Fig.~\ref{fig:def_geometry}a the solution must be
regular for $\Theta\le\theta\le\pi$. Instead of using combinations of $P$ and $Q$,
we can simply use $P_{\eta_0}(-\cos\theta)$, which will be regular at $\cos\theta=-1$.
The value of $\eta_0$ is then determined by requiring
\begin{equation}
\label{Eq:eigen_cone}
P_{\eta_0}(-\cos\Theta)=0.
\end{equation}
Since $\Psi$ cannot change sign in the physically permitted region, the
smallest possible $\eta_0$ must be chosen. The corresponding solution has been described
in detail in Ref.~\cite{BenNaim}. In particular, in the limit
$\Theta\to0$, the exponent $\eta_0$ vanishes through a logarithmic
singularity ($\sim1/|\ln\Theta|)$, while for $\Theta\to\pi$ it diverges as
$\eta_0\approx2.405/(\pi-\Theta)$.

For the geometry of a polymer attached to the apex of a cone touching an infinite
plane (Fig.~\ref{fig:def_geometry}c), we need to solve Eq.~\eqref{Eq:Legendre3D}
with $\Psi(\theta)$ vanishing  at both $\Theta$ and $\pi/2$. The latter
condition corresponds to $\mu=0$ and can be
assured by setting in Eq.~\eqref{Eq:3D_solution} the ratio
\begin{equation}
\frac{a_2}{a_1}=-\frac{P_{\eta_0}(0)}{Q_{\eta_0}(0)}=\frac{2}{\pi\tan(\pi\eta_0/2)}.
\end{equation}
With the above choice, the angular function can be conveniently written as
\begin{equation}
\label{Eq:antisympsi}
\Psi(\theta)=\pi\sin(\pi\eta_0/2)P_{\eta_0}(\cos\theta)
+2\cos(\pi\eta_0/2)Q_{\eta_0}(\cos\theta).
\end{equation}
For non-integer $\alpha$, $P_\alpha(\mu)$ and $P_\alpha(-\mu)$ are
linearly independent and both
solve Eq.~\eqref{Eq:Legendre3D}~\cite{NISTlib}. Thus for non-integer $\eta_0$, Eq.~\eqref{Eq:antisympsi} can be replaced by

\begin{equation}
\label{Eq:simplified}
\Psi(\theta)=P_{\eta_0}(\cos\theta)-P_{\eta_0}(-\cos\theta).
\end{equation}
Again, we must choose the smallest $\eta_0$ for which this function vanishes
for $\theta=\Theta$. The resulting exponents (which cannot be cast as simple functions),
are plotted as solid lines in Fig.~\ref{fig:ThAll}.
From  bounds on the roots of Legendre polynomials~\cite{Szego36,markoff,stieltjes},
we find that in the limit $\Theta\to\pi/2$, the exponent diverges as
$\eta_0\approx 2\pi/(\pi-2\Theta)$, while as $\Theta\to0$, $\eta_0\to1$.
Indeed, the solution in Eq.~\eqref{Eq:antisympsi} has no roots for $\eta_0<1$.
Asymptotic expressions for $P_\alpha(\mu)$ and $Q_\alpha(\mu)$ near $\mu=1$ are known and can be
used to determine  positions of roots in such limit, indicating that
$\eta_0\approx 1+1/|\ln(\Theta)|$.

It is worth noting that in all cases above ($d=2,3,4$)
the cone+plate exponent has the {\it identical} divergence
$\eta_0\approx2\pi/(\pi-2\Theta)$ as $\Theta\to\pi/2$. This may be justified by arguing that a
diffuser confined between an almost flat cone and the plate encounters
absorbing boundaries not dissimilar to that of a wedge.

\par\noindent$\bullet$
{\it In general $d$,} with the substitution
$\Psi(\theta)=\sin^{-\delta}\theta\ u(\mu)$, where $\mu=\cos\theta$ and
$\delta=(d-3)/2$, Eq.~\eqref{Eq:psi} becomes
\begin{align}
\label{Eq:GenLegender}
&(1-\mu^2)\frac{d^2u}{d\mu^2}-2\mu\,\frac{du}{d\mu}+       \nonumber \\
&\left[(\eta_0+\delta)(\eta_0+\delta+1)-\frac{\delta^2}{1-\mu^2}\right]u=0,
\end{align}
which is solved by associated Legendre functions~\cite{NISTlib}
that can be chosen in several forms such as
$P_{\eta_0+\delta}^{\pm \delta} (\pm \mu)$,
$Q_{\eta_0+\delta}^{\pm \delta} (\pm \mu)$, or their combinations.
The particular choice of one or pair of linearly independent functions
depends on the boundary conditions and the specific values of $\eta_0$
and $\delta$. For the case of cone in different dimensions the reader is
referred to Ref.~\cite{BenNaim}. To find a proper form for $u$ in (integer
or fractional) $d$ dimensions, we can closely follow the case of $d=3$:
For a  single cone we may chose
$u=P^{-\delta}_{\eta_0+\delta}(-\cos \theta)$,
since $\sin^{-\delta}\theta\ P^{-\delta}_{\eta_0+\delta}(-\cos \theta)$
has no cusp at $\theta=\pi$. The value of $\eta_0$ will be set by the
requirement that the function vanishes for $\theta=\Theta$. For cone+plate
configurations we may choose
\begin{equation}
\label{Eq:plateGeneralD}
u=a_1P^{-\delta}_{\eta_0+\delta}(-\cos \theta)-a_2P^{-\delta}_{\eta_0+\delta}(\cos \theta)\,,
\end{equation}
which for $a_1=a_2$ vanishes on the plate ($\theta=\pi$), and
will be a suitable solution as long as $\eta_0$ is not integer and is chosen
such that the function vanishes for $\theta=\Theta$.

{\it For all $d$,} the cone becomes a plate for $\Theta=\pi/2$.
Correlations with one point approaching a surface are easily obtained by
the method of images~\cite{chandra,Binder83} leading to
$\eta_{\rm c{0}}=\eta_{\rm p{0}}=1$, which is clearly seen from the
intersection of the curves in Fig.~\ref{fig:ThAll}.

Note that in $d=3$ both cone and cone+plate exponents  approach
their limit $\Theta\to 0$ via a logarithmic singularity, while
for $d=4$ they approach that limit linearly. For intermediate dimensions
$3<d<4$, the limiting behavior as $\Theta\to 0$ is given by
\begin{align}\label{Eq. eta0 needle}
  \eta_{\rm c0} &= \frac{\Gamma(1-\epsilon/2)}{\sqrt{\pi}\, \Gamma(1/2-\epsilon/2)}\Theta^{1-\epsilon}, \qquad{\rm and} \nonumber \\
  \eta_{\rm cp0} &= 1+\frac{4\Gamma(2-\epsilon/2)}{\sqrt{\pi}\, \Gamma(1/2-\epsilon/2)}\Theta^{1-\epsilon},
\end{align}
where $d=4-\epsilon$.
The singular behavior in the above equations is a power law $\Theta^{p_0}$ with $p_0=d-3$,
which has a simple geometric interpretation.
The ideal polymer can be regarded as a self-similar object with fractal dimension 2,
while the remnant of the cone as $\Theta\to0$ is a (semi-infinite) 1-dimensional line.
When embedded in $d$-dimensional space, the intersection of the two
entities (random walk and remnant line) spans a space of dimension $d-(2+1)=p_0$.
Indeed, for $d<3$ the limiting value is different from the case without any
cone indicating the finite probability of intersection of the polymer with the semi-infinite barrier line.

In both $d=2$ and $d=4$, we found simple expressions for the ideal
polymer attached to the contact point between two coaxial cones of
opening angles $\Theta_1$ and $\Theta_2$. However, in general $d$
the solution has the form in Eq.~\eqref{Eq:plateGeneralD}.
The vanishing of $u$ on the first cone requires
$a_1/a_2=\cos(\Theta_1)/\cos(-\Theta_1)$. With this choice, the
only free parameter left is the exponent $\eta_0$.
To satisfy the condition $u(\theta=\Theta_2)=0$, we must have
\begin{align}
  P^{-\delta}_{\eta_0+\delta}(-\cos \Theta_1)  &P^{-\delta}_{\eta_0+\delta}(-\cos \Theta_2)-\nonumber \\
  &P^{-\delta}_{\eta_0+\delta}(\cos \Theta_1) P^{-\delta}_{\eta_0+\delta}(\cos \Theta_2)=0.
\end{align}
Unfortunately, values of $\eta_0$ satisfying the above equation cannot be expressed
as simple functions. However,
for small $\Theta_1$ and $\Theta_2$, we can prove that
\begin{equation}\label{Eq. Touching cones}
\eta_{\rm cc0}(\Theta_1,\Theta_2)\approx\eta_{\rm c0}(\Theta_1)
+\eta_{\rm c0}(\Theta_2),
\end{equation}
i.e. the exponent for two sharp cones is approximately the sum of exponents for individual
cones of opening angles $\Theta_1$ and $\Theta_2$.

\section{Self-avoiding polymers:\\ Simulations}\label{sec:SAW}
Self-avoiding walks (SAWs) provide a convenient model for exploring
universal aspects of swollen (coil) polymers with short-range interactions.
The number of SAWs in three-dimensional space (without obstacles) is governed by
the critical exponent $\gamma\approx1.158$~\cite{caracciolo} (corresponding to
$\eta\approx 0.03$),
while in two dimensions, $\gamma=43/32$~\cite{madras_book} ($\eta=5/24$). (For ideal
polymers these exponents do not depend on $d$:  $\gamma_0=1$ and $\eta_0=0$.)
Several important results regarding $\gamma_{\rm s}$ (with obstacles)
for polymers confined by wedges or planes are
known~\cite{cardy_red,cardy1,cardy2,guttmann,gsurface,debell}.
For example the exponent $\gamma_{\rm wedge}$ of a SAW confined to
a wedge (in $d=2$ or $d=3$), and anchored at its sharp end,
depends on opening angle and
diverges to $-\infty$ as the confining angle vanishes. Polymers
attached to the tip of a two-dimensional sector in $d=3$, and to the
apex of a cone also were studied~\cite{slutsky}. Numerous analytical
\cite{kosmas,Douglas89} and numerical~\cite{gsurface,debell}
studies of SAWs anchored to flat surfaces in $d=3$ find
$\gamma_{\rm s} \equiv \gamma_{\rm p}$ in the range of 0.70
\cite{gsurface} to 0.68~\cite{grass_gamma1}
($\eta_{\rm p}=0.81$ or 0.84, respectively). (Again, for ideal polymers
these exponents do not depend on $d$:
$\gamma_{\rm p0}=1/2$ and $\eta_{\rm p0}=1$.)
We are particularly interested in geometries depicted in
Fig.~\ref{fig:def_geometry}, and provide below our numerical and
analytical estimates of the relevant exponents.

We employed a dimerization method~\cite{dimerization1,dimerization2} to
numerically generate  SAWs on a cubic lattice. In this (recursive) method two
$N/2$-step SAWs are joined in an attempt to create an $N$-step walk.
If the resulting walk does not self-intersect, the process is
successful and the $N$-step walk is used to generate even larger SAWs.
If the combined walk has intersections, it is rejected and both $N/2$-step
components are discarded. The process is then repeated. As the rejection rate
increases slowly with $N$, this is an extremely efficient method for
creating an unbiased collection of SAWs. We generated $10^8$ SAWs of
lengths $N=16,~32,~\cdots,~1024$. Each SAW was attached
to the origin, and we checked whether it intersects the imposed obstacles.

\begin{figure}
\null\vskip 1cm
\includegraphics[width=8cm]{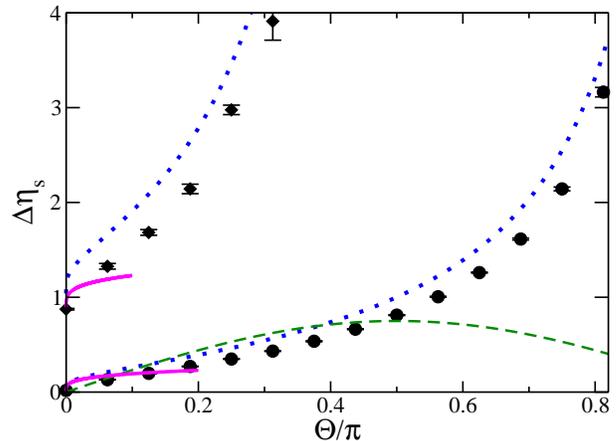}
\caption{
\label{fig:detavstheta}
(Color online) Dependence of the exponent difference
$\Delta\eta_s=\eta_{\rm s}-\eta$ on apex semi-angle $\Theta$
for the  cone+plate (top curves), and an isolated cone (bottom curves).
The dotted curves are the exact values for an {\it ideal} polymer. The full
diamonds and circles represent numerical results for SAWs in the same geometry.
Error bars show the uncertainty in $N\to\infty$ extrapolation.
The dashed line depicts the $\epsilon$-expansion result in Ref.~\cite{slutsky}
(with $\epsilon=1$) for `weak surface repulsion.'
The solid lines show the values of $\eta_{\rm c}$ from Eq.~\eqref{Eq:eta c}, and
$\eta_{\rm cp}$ from Eq.~\eqref{Eq:eta_cp}, with $\epsilon=1$. For the latter,
$\Delta\eta_s=\eta_{\rm s}$ since $\eta=0+O(\epsilon^2)$ in free space.
}
\end{figure}

The probability that a SAW does not intersect the confining boundaries
is the ratio of the number of SAWs satisfying the geometrical constraints
to the total number of such walks, i.e.
$p_N={\cal N}_{\rm s}/{\cal N}\sim N^{\gamma_{\rm s}-1}/N^{\gamma-1}
=N^{\Delta\gamma_{\rm s}}$, and therefore
$p_N/p_{2N}=2^{\Delta\gamma_{\rm s}}$. This result becomes accurate
for $N\to\infty$, and we extract the limiting value of $\gamma_{\rm s}$
by plotting its finite-$N$ estimates as a function of $1/\sqrt{N}$.
The estimated errors in the extrapolated values of the exponents
are caused both by the limited sample size, and by values of $N$ that are not long enough.
The resulting exponents
$\Delta\eta_{\rm s}\equiv \eta_{\rm s}-\eta=(\gamma-\gamma_{\rm s})/\nu$
(full symbols in Fig.~\ref{fig:detavstheta}) are somewhat lower than
the corresponding values for ideal polymers (indicated by dotted lines).
The difference primarily originates in the change in $\eta_{\rm s}$
as a result of self-avoiding interactions. (This trend is slightly moderated
by a small shift of $\eta$ in  free space.) Figure~3 of
Ref.~\cite{MKK_EPL96} presents $\Delta\gamma_{\rm s}$ as a function of
$\Theta$, in which form results for SAWs are much closer to those
of ideal polymers due to multiplication by the self-avoiding and ideal
values of $\nu$, respectively.

The dashed line in Fig.~\ref{fig:detavstheta} represents the result of
an $\epsilon=4-d$ expansion~\cite{slutsky} that treats both the self-repulsion of
the polymer, as well as its repulsion by the two-dimensional {\em surface of the cone},
as  weak perturbations. A renormalization group computation is then
carried out to the lowest order in $\epsilon$, resulting in
$\Delta\eta_{\rm c}=(3\epsilon/4)\sin\Theta$. For
small $\Theta$ this expression resembles the expected behavior, but for
$\Theta>\pi/2$ the function (incorrectly) decreases.
This is because, faced with the weakly repulsive potential at the surface of the cone,
 the polymer simply jumps from the decreasing
exterior into
the larger internal space of the cone. This approach also produces an
incorrect $\Theta\to0$ behavior. Deficiencies of the results of
Ref.~\cite{slutsky} can  be remedied by exclusion of the entire
{\em interior of the cone}, as was done by Cardy~\cite{cardy1,cardy2}
for the wedge geometry. The analogous computations for a cone, described
in the following section,  are  more complicated, and build upon more
recent results pertaining to the electrodynamic Casimir interactions between
conducting cones and plates~\cite{Maghrebi10}.

\section{Self-avoiding polymers:\\ Epsilon expansion}\label{sec:epsilon}
\subsection{Cone}

The constraint of self-avoidance is relevant in dimensions $d\leq 4$, where exponents
are computed perturbatively in $\epsilon=4-d$.
A primary element of such calculation is the  Green's function in the absence of interactions
(self-avoidance), but in the presence of obstacles.
For the latter we need the full solution to the Laplace equation in
four dimensions; {therefore we should go beyond Eq.~(\ref{Eq:psi}) which is only applicable at large separations}. {In general, one should find the Green's function in $4-\epsilon$ dimensions. However, we are interested in finding corrections to the first order in $\epsilon$. With the strength of the interaction being of the same order, we can safely limit ourselves to computing the Green's function in four dimensions.}
Subject to appropriate boundary conditions, the Green's function is the solution to
\begin{equation}
  -{\nabla^2_{(4)}} G_0(x, x')=\delta^{4}(x -x'),
\end{equation}
where $x$ and $x'$ are spatial coordinates, and the subscript $4$ emphasizes the space dimension.
Since we are interested in conical boundaries, it is convenient to break up the Green's function along the polar angular coordinate.
In order to accomplish this, we need a complete set of functions of the other three coordinates including the radius. One can exploit the Kontrovich-Lebedev transform~\cite{Samko1993} to formulate the completeness relation. The procedure is analogous to Ref.~\cite{Maghrebi10} where some of the authors have carried out a similar analysis.
Details of the calculation are given in the Appendix; the final result for the
 Green's function for a single cone is
\begin{widetext}
\begin{align}
\!\!G_0(x,x')= \!\! {\sum_{nlm} }\frac{ (-1)^l\pi\Gamma(\rho_n+l+1)}{2\sin(\rho_n \pi) \Gamma(\rho_n-l)}
\frac{P_{\rho_n-1/2}^{-l-1/2}(\cos\Theta)}{\partial_{\rho_n}P_{\rho_n-1/2}^{-l-1/2}(-\cos\Theta)}
\frac{r_<^{\rho_n-1}}{r_>^{\rho_n+1}}
\frac{P_{\rho_n-1/2}^{-l-1/2}(-\cos\theta)Y_{lm}(\psi, \phi)}{\sqrt{\sin\theta}}
\frac{P_{\rho_n-1/2}^{-l-1/2}(-\cos\theta')Y^{\star}_{lm}(\psi', \phi')}{\sqrt{\sin\theta'}}\,
,\quad
\label{Eq:G0}
\end{align}
\end{widetext}
where the summation is made over a triplet of integers $n>0,~l\geq 0$, and $-l\leq m \leq +l$.
The exponent $\rho_n$ labeled by the integer $n$ is the root of the equation
\begin{equation}
\label{Eq:rhok}
  P_{\rho_n-1/2}^{-l-1/2}(-\cos\Theta)=0.
\end{equation}
In the above equation, $r^<$ and $r^>$ refer to the smaller and larger radial coordinate
for $x$ and $x'$. Such division of the Green's function is appropriate since one endpoint
of the polymer is close to the tip of the cone and the other end is far away.

For an interacting Green's function, one should subtract polymer configurations
which self-intersect.
In the perturbative analysis, the strength of self-repulsion is indicated by a parameter $u$,
and to lowest order one must subtract contributions forming a single intermediate loop, such that
\begin{equation}
G_1=
G_0
\! -u\!\!\int \!{\rm d}^4{ x''}\! G_0(x,x'')
G_0^r(x''\!,x'')  G_0(x''\!,x').
\label{Eq:G1}
\end{equation}
Similar expansions arise in the context of quantum field theories, in which language the first term in the last equation is the ``free'' propagator (in the presence of external boundary conditions) while the next term is the ``one-loop'' correction. The latter should be regularized by replacing the middle Green's function (computed at identical points) with the difference between the Green's functions in the presence and absence of external boundaries, {i.e.} $G_0^r=G_0- \bar G_0$.
In a renormalization group treatment, the parameter $u$ changes with scale of observation,
ultimately arriving at a fixed point value of $u^*=2\pi^2\epsilon$~\cite{slutsky,cardy1,cardy2}.
This universal value is a characteristic of the polymer and independent of obstacles or
boundary conditions.
To find the scaling behavior for $r\gg a$, it is sufficient to include only
the term $n=1,~l=m=0$ from the sum in Eq.~\eqref{Eq:G0} in the non--loop propagators, arriving at
\begin{align}\label{Eq. G0 c: leading}
  G_0{\large|}^{n=1}_{l,m=0}(x,x')= \frac{r_<^{\rho_1-1}}{{4\pi^2} r_>^{\rho_1+1}} \frac{\sin \rho_1\,(\pi-\theta)}{\sin\theta}\frac{\sin \rho_1\,(\pi-\theta')}{\sin\theta'}.
\end{align}
(Corrections from higher-order terms are in higher powers of $a/r$, similar to the analysis by Cardy ~\cite{cardy1,cardy2}.)
However, the loop (Green's function at identical points) can be of small size; in fact, we have regularized this Green's function to cancel out the divergent contribution in the limit of a shrinking loop. Therefore, for the loop propagator one should consider the entire sum.

The integral corresponding to the loop variable is over the whole space:  The azimuthal and spherical angles $\phi''$ and $\psi''$ are trivial due to rotational symmetry.
Once summed over spherical harmonics indexed by the same $l$, we get a factor of $l(l+1)$. The integral over the radial coordinate, $r''$, should produce a logarithm---later to be exponentiated to a power-law.
Along the line of Cardy's analysis,
this logarithm is due to the contribution from the region $r<r''<r'$.
Finally, an integration over $\theta''$, the polar angle, completes the integral; however, the latter is complicated by the fact that the roots of the non-algebraic Eq.~(\ref{Eq:rhok}) should be computed.

Considering the above complications, we focus instead on a cone with small opening angle, i.e. $\Theta\to 0$. In this limit, one can see that the leading
singularity  is due to the $l=0$ term in the Green's function corresponding to the loop. Note that a sharp cone is invisible to higher spherical partial waves. The loop Green's function is then obtained from the sum
\begin{align}\label{Eq. small angle Green's function}
  G_0^r{\large|}_{l=m=0}(x{''},x{''})= \frac{1}{4\pi^2 {r''}^2} \sum_{n=1}^{\infty} \frac{1}{n} \frac{\sin^2n\pi\frac{\pi-\theta''}{\pi-\theta_0}-\sin^2n\theta''}{\sin^2\theta''}.
\end{align}
Equations~(\ref{Eq. G0 c: leading}) and~(\ref{Eq. small angle Green's function}) are used to  perform the integral over $\theta''$ and summation over all $n$. The Green's function dependence on the radius is then obtained as
\begin{equation}\label{Eq:DoubleLog}
G_1\propto \frac{r_<^{\eta_{\rm c 0}}}{{r_>}^{\eta_{\rm c0} +2}}\left[1+ \epsilon\ln\frac{r_<}{r_>}\left(\frac{1}{4\pi}\Theta\ln\Theta+.16\, \Theta\right)\right],
\end{equation}
where the second term in the bracket is the correction due to the loop integral, and thus it is proportional to $\epsilon$. Another dependence on $\epsilon$ is introduced through the exponent $\eta_{\rm c0}$ whose value in $d=4-\epsilon$ dimensions is given by Eq.~\eqref{Eq. eta0 needle}. The radial logarithm can be exponentiated yielding the renormalized exponent,
\begin{equation}\label{Eq:Deta1}
  \eta_{\rm c}=\eta_{\rm c0} + \left(\frac{\Theta\ln\Theta}{4\pi}+.16 \, \Theta\right) \, \epsilon\,.
\end{equation}
As the first correction to $\eta$ in empty space appears at the order of $\epsilon^2$,
 the above (first-order) result for the cone vanishes logarithmically as  $\Theta\to0$.
The logarithm in the angular variable is suggestive of another exponentiation to obtain a power-law in the limit of $\Theta\to0$.
First note, however, that expanding {$\eta_{\rm c0}$ to the first order in $\epsilon$ yields
\begin{equation}
  \eta_{\rm c0} \approx \frac{\Theta}{\pi}+\left(-\frac{\Theta\ln\Theta}{\pi} -.22 \, \Theta\right)\,\epsilon \,,
\end{equation}
producing another contribution to $\ln\Theta$, originating
from the expansion in $4-\epsilon$ dimensions of the phantom polymer (as
opposed to the perturbative terms from the one-loop computation).
Putting all these pieces together, we obtain
\begin{align}
  \eta_{\rm c}= \frac{\Theta}{\pi}\left(1-\frac{3}{4}\,\epsilon\ln\Theta-.06\,\epsilon\right),
\end{align}
which can be recast as a power law
\begin{equation} \label{Eq:eta c}
\eta_{\rm c}\approx\left(\frac{1}{\pi}-.06\,\epsilon\right)\Theta^{1-\frac{3}{4}\epsilon} \,.
\end{equation}
As expressed in Eq.~\eqref{Eq. eta0 needle}, the exponent $\eta_{\rm c0}$ vanishes
with cone angle as $\Theta^{p_0}$ with $p_0=d-3=1-\epsilon$ for the phantom polymer (Fig.~\ref{fig:ThAll}). The above equation indicates that the vanishing of $\eta_{\rm c}$ for a
self-avoiding polymer is governed by the modified exponent $p=1-3\epsilon/4$.
This dependence is shown for $\epsilon=1$ as the lower solid line in
Fig.~\ref{fig:detavstheta}. We may interpret this result as follows:
The self-avoiding condition swells the polymer at all scales compared to a random walk.
As a result, the fractal dimension of the polymer is reduced from 2
to $\nu^{-1}=2- \epsilon/4$ at the linear order in $\epsilon$.
Whether or not a  fractal coil and the one-dimensional needle  intersect depends
on the dimensionality  $d$ of space; the domain of their intersection is given
by the co-dimension of the polymer+needle, i.e.
$d-1-\nu^{-1}=1-3\epsilon/4=p$.
It would be interesting to see if this connection holds in higher orders in perturbation theory.
The current numerical results are not accurate enough to test this conjecture.

Finally, we note that for two touching cones of small angles $\Theta_1$ and $\Theta_2$, we find the analog of Eq.~(\ref{Eq. Touching cones}) to first order in the $\epsilon$-expansion, namely
\begin{equation}
\eta_{\rm cc}(\Theta_1,\Theta_2)\approx\eta_{\rm c}(\Theta_1)+\eta_{\rm c}(\Theta_2).
\end{equation}

\subsection{Cone-Plate}
When constrained by a cone touching a plate, the Green's function is given by
\begin{widetext}
\begin{align}
&\!\!G_0(x,x')= {\sum_{nlm} } \frac{(-1)^l\pi}{2}  \frac{r_<^{\rho_n-1}}{r_>^{\rho_n+1}} \frac{\Gamma(\rho_n+l+1)}{\sin(\rho_n \pi) \Gamma(\rho_n-l)} \frac{P_{\rho_n-1/2}^{-l-1/2}(\cos\Theta)}{\partial_{\rho_n}\left(P_{\rho_n-1/2}^{-l-1/2}(-\cos\Theta)-P_{\rho_n-1/2}^{-l-1/2}(\cos\Theta)\right)} \times
\nonumber \\
&\frac{(P_{\rho_n-1/2}^{-l-1/2}(-\cos\theta)-P_{\rho_n-1/2}^{-l-1/2}(\cos\theta))Y_{lm}(\psi, \phi)}{\sqrt{\sin\theta}} \frac{(P_{\rho_n-1/2}^{-l-1/2}(-\cos\theta')-P_{\rho_n-1/2}^{-l-1/2}(\cos\theta'))Y^{\star}_{lm}(\psi', \phi')}{\sqrt{\sin\theta'}}\,,
\end{align}
where the sum is again over the triplet of integers $n>0,~l\geq 0$, and $-l\leq m \leq +l$.
The exponent $\rho_n$ is now the $n$-th root of the transcendental equation
\begin{equation}
  P_{\rho_n-1/2}^{-l-1/2}(-\cos\Theta)-  P_{\rho_n-1/2}^{-l-1/2}(\cos\Theta)=0.
\end{equation}
Again for the non-loop propagators, it suffices to keep only the first term in the sum
\begin{align}\label{Eq. G0 cp: leading}
  G_0{\large|}^{n=1}_{l,m=0}(x,x')= \frac{r_<^{\rho_1-1}}{{4\pi^2} r_>^{\rho_1+1}} &\frac{\sin \rho_1\,(\pi/2-\theta)}{\sin\theta}\frac{\sin \rho_1\,(\pi/2-\theta')}{\sin\theta'}\,.
\end{align}
However, the intermediate loop should be summed entirely. Here again, we focus on a sharp cone. It is more convenient to express the regularized Green's function for the intermediate loop as
\begin{equation}\label{Eq: Gp}
  G^r_0\equiv G_0-\bar G_0= (G_0-G^p_0)+(G^p_0-\bar G_0)\,,
\end{equation}
where $G^p_0$ is the Green's function in the presence of the plate alone and is independent of the cone angle $\Theta$. Also note that $\bar G_0$ is the Green's function in empty space. Each bracket on the RHS of Eq.~(\ref{Eq: Gp}) can be treated separately. The first bracket, to the leading order for a sharp cone, is given by
\begin{align}
  \left(G_0-G^p_{0}\right){\large|}_{l=m=0}(x'',x'')= \frac{1}{4\pi^2 {r''}^2} \sum_{n=1}^{\infty} \frac{1}{n}  \frac{\sin^2n\pi\frac{\pi/2-\theta''}{\pi/2-\theta_0}-\sin^2 2n\theta''}{\sin^2\theta''}\,.
\end{align}
Also the second bracket in Eq.~(\ref{Eq: Gp}) is merely the Green's function in the presence of a plate. The latter can be cast as a series expansion by using the method of images:
\begin{align} \label{plate G-G0}
&\left(  G^p_0-\bar G_0\right)(x,x')= - {\sum_{nlm} }\frac{r_<^{n-1}}{r_>^{n+1}} \frac{\Gamma(n+l+1)}{2\Gamma(n-l)} P_{n-1/2}^{-l-1/2}(-\cos\theta)P_{n-1/2}^{-l-1/2}(\cos\theta') \, Y_{lm}(\psi, \phi) Y^{\star}_{lm}(\psi', \phi').
\end{align}
\end{widetext}
The Green's functions appropriate to the external legs and the loop should be inserted in Eq.~(\ref{Eq:G1}) to obtain, in the limit of  $\Theta\to0$,
\begin{equation}
G_1\propto \frac{r_<^{\eta_{\rm cp 0}}}{{r_>}^{\eta_{\rm cp0} +2}}\left[1+ \epsilon\ln\frac{r_<}{r_>}\left(-\frac{1}{8}+\frac{\Theta\ln\Theta}{\pi} +.66\,\Theta\right)\right],
\end{equation}
where the exponent $\eta_{\rm cp0}$ is given by Eq.~\eqref{Eq. eta0 needle}. Upon exponentiation in $r$, we find
\begin{eqnarray}\label{Eq:Deta2}
\eta_{\rm cp}=\eta_{{\rm cp}0}+\left(-\frac{1}{8}+\frac{\Theta\ln\Theta}{\pi} +.66\,\Theta\right)\, \epsilon \,.
\end{eqnarray}
Note that the loop-correction does not vanish for $\Theta\to0$, instead going to the limiting
value of $-\epsilon/8$ due to the presence of the plate.
Expanding $\eta_{\rm cp 0}$ to the first order in $\epsilon$,
\begin{equation}
   \eta_{\rm cp0}(\Theta, \epsilon) \approx 1+\frac{4\Theta}{\pi}+\left(-\frac{4\Theta\ln\Theta}{\pi} -1.52 \, \Theta\right)\,\epsilon\,,
\end{equation}
we find a logarithmic dependence in $\Theta$. Summing both contributions, the exponent $\eta$ to the lowest order in $\epsilon$ and $\Theta$ becomes
\begin{align}
  \eta_{\rm cp}&= 1-\frac{\epsilon}{8}+\frac{4\Theta}{\pi}\left(1-\frac{3}{4}\,\epsilon\ln\Theta-.86\,\epsilon\right).
\end{align}
As in the previous section, we  exponentiate the logarithm in $\Theta$ to obtain the power-law dependence
\begin{equation}\label{Eq:eta_cp}
\eta_{\rm cp} \approx1-\frac{\epsilon}{8}+
\left(\frac{4}{\pi}-.86\,\epsilon\right)\Theta^{1-\frac{3}{4}\epsilon}.
\end{equation}
Revealingly, the exponent $p=1-3\epsilon/4$ is the same as in the case of a single cone, whereas
the amplitude has changed. Furthermore, it approaches a constant value in the
limit of vanishing angle. This result is shown for $\epsilon=1$ as the
upper solid line in Fig.~\ref{fig:detavstheta}.

Finally we can compute the amplitude of the force according to Eq.~(\ref{Eq:A}), using Eqs.~(\ref{Eq:eta c}) and~(\ref{Eq:eta_cp}), as
\begin{equation}\label{Eq: force}
  {\cal A}= 1-\frac{\epsilon}{8}+ \left(\frac{3}{\pi}-.80\,\epsilon\right)\Theta^{1-\frac{3}{4}\epsilon}.
\end{equation}

\subsection{Star Polymers}

In a typical setup, the amplitude ${\cal A}=\eta_{\rm cp}-\eta_{\rm c}$ with the exponents being computed in the previous (sub)sections is of order unity. Therefore, according to Eq.~\eqref{Eq:force}, the force is roughly 0.1 pN at room temperature  in a separation of 0.1 $\mu$m. Such a force is at the margin of measurement by current precision apparatus.
We can further increase the force by attaching more polymers to the cone tip. The total force is additive for ideal (phantom) polymers, i.e. it is proportional to $f$, the number of arms. For self-avoiding polymers, however, interactions come into play and the result is no longer additive.

In general, there are two types of interactions: a single arm can self-intersect
(intra-arm interaction), or two different arms intersect (inter-arm interaction).
The  former effect leads to corrections similar to those computed in previous
(sub)sections, and would by itself simply lead to
multiplication of the force on a single polymer by $f$.
The latter interaction, originating from intersections between two different arms,
is proportional to $f(f-1)/2$, the number of interacting pairs. We can then write
the overall exponent as  $\eta=f\eta_0+f\eta^i+\frac{f(f-1)}{2}\eta^e$, where $\eta_0$ characterizes
a single phantom polymer, and $\eta^i$ and $\eta^e$ correspond to intra- and
inter-arm interactions respectively.   A general situation of many-arm polymers
in the absence of external boundaries is considered in detail in Ref.~\cite{Ohno89}.

To zeroth order (i.e. in the absence of interactions) the correlation function describing a
pair of polymers starting at the same point $x$, but with different endpoints at $x'_{1}$
and $x'_{2}$, is simply the product of two free Green's functions, $G_0 \otimes G_0$.
Subtracting configurations in which the two polymers avoid each other yields to first order
 the interacting Green's function
\begin{align}\label{Eq: star}
{G^{(2)}_1}'=G_0 \, G_0-u\int {\rm d}^4x''G^2_0(x,x'') G_0(x'', x'_{1}) G_0(x'', x'_{2})\,,
\end{align}
where we have integrated over the intersection point $x''$.
The prime on the Green's function indicates that we have only considered the inter-arm interactions. Also note that no regularization is needed since the integral in Eq.~(\ref{Eq: star}) is finite. For a long polymer, it is sufficient to include only the first term in the Green's function, i.e. Eq.~(\ref{Eq. G0 c: leading}) for a single cone and Eq.~(\ref{Eq. G0 cp: leading}) for a cone-plate configuration. The leading logarithmic contribution for a cone is then
\begin{equation}
  {G^{(2)}_1}'= G_0 \, G_0\left(1+\epsilon \ln \frac{r}{r_{\rm max}} \left(\frac{1}{4}-\frac{\Theta}{2\pi}\right)\right),
\end{equation}
where $r_{\rm max}$ depends only on $r'_1$ and $r'_2$. Hence, by exponentiating the radial function, $\eta^e$ for a cone is obtained as
\begin{equation}
  \eta_{\,{\rm c}}^e=\epsilon \left(\frac{1}{4}-\frac{\Theta}{2\pi}\right).
\end{equation}
The corresponding $\epsilon$-expansion for cone+plate gives
\begin{equation}
  \eta_{\,{\rm cp}}^e=\epsilon \left(\frac{1}{4}-\frac{7\Theta}{3\pi}\right).
\end{equation}
These equations then lead to the force amplitude
\begin{equation}
\label{starA}
\frac{{\cal A}(f)}{f}=1-\frac{\epsilon}{8}+
\left[\frac{3}{\pi}-\left(.80+\frac{11}{12\pi}(f-1)\right)\epsilon\right]\Theta^{1-3\epsilon/4}\,.
\end{equation}
Interestingly, inter-arm interactions  {\it reduce} the force amplitude per polymer.
This equation is similar to Eq.~(\ref{Eq: force}) with the addition of inter-arm interactions.

Interestingly, the same exponent $p=1-3\epsilon/4$ dictates the limiting behavior in Eq.~(\ref{starA}) as $\Theta\to0$.
Yet another possible experimental set-up is a system consisting of a polymer attached to a cone which is approaching
another cone. The entropic force is relatively
smaller in this case. We have verified that the exponent in the latter case also vanishes with the opening angle
as $\Theta^p$, providing further support for the conjecture that this exponent reflects
the intersection of a fractal polymer and a needle.

\section{Discussion}

In summary, we have demonstrated that polymers exert an entropic force ${\cal A}k_BT/h$
on a cone tip, with a `universal' amplitude $\cal A$ dependent on geometry,
interactions, and internal topology of the polymer. We  conjecture that the singular
behavior of the amplitude on vanishing cone angle is described by a new
exponent, simply related to the fractal dimension of the polymer.
There are many  other self-similar shapes  where a similar force law is expected on the basis
of scaling at length scales shorter than an appropriate correlation length.

In this work we concentrated on situations where the polymer is attached to
a single surface such as described by Fig.~\ref{fig:def_geometry}b or
\ref{fig:def_geometry}d. However, our results can be easily applied to  a variety of
other situations. First, we note that our approach is easily generalized to a
slightly more complicated situation with one polymer attached to a cone while
another is attached to a surface: The resulting force coefficient ${\cal A}$ is
 given by Eq.~\eqref{Eq:A} with $\eta^{\rm initial}$ consisting of the {\em sum}
of $\eta_{\rm c}$ for the two cones, while $\eta^{\rm final}$ corresponds to
two polymers (or a star polymer with $f=2$) attached to the contact point
between a cone and a plate (``s=cp"). An even more relevant situation is the
force constant of a {\em single} polymer attached at both ends to a cone and
a plate as in  Fig.~\ref{fig:def_geometry}e. One may view this situation as
an elaboration of the  previous  case of two polymers, when their free ends
are joined to each other. Such a modification significantly changes the
behavior of the system when $h$ is of order or larger than $R_0$. In fact the
force changes sign at $h\sim R_0$, when the polymer state changes from
stretched to compressed. For $h\ll R_0$ connecting the polymer ends modifies
the free energy of the system. However, it is plausible that in such limit
the force transmitted by the polymer(s) is not influenced by end-point
connection, and therefore the force amplitude may be the same as the case
of two disconnected polymers. Verification of this point requires a further
study of the forces between the polymer and the boundaries.

\begin{acknowledgments}
This work was supported by the National Science Foundation under Grants
No.~DMR-12-06323 (MK, MFM), and  PHY05-51164 (MK at KITP). YK acknowledges
the support of Israel Science Foundation grant 99/08.
The authors acknowledge discussions with B. Duplantier and A. Grosberg.
\end{acknowledgments}

\appendix
\section{Green's functions}\label{Appendix}
We first derive the Green's function for the Helmholtz equation in four dimensions;
the corresponding function for the Laplace equation is then obtained as a  limit.
The former satisfies the Helmholtz equation with a delta function source:
\begin{equation}\label{Eq. Helmholtz eqn with 4 coordinates}
  -\left({\nabla^2_{(4)}}+k^2\right)G_0(x, x')=\delta^{4}(x -x')\,;
\end{equation}
the latter is obtained by setting the wavevector $k$ to zero.
Consider spherical coordinates $x=(r, \theta, \psi, \phi)$, where $r$ is the radius, $\theta$ is the polar angle and $\phi$ and $\psi$ the remaining angular coordinates---a constant $\theta$ specifies the surface of a cone in four dimensions. The cone is {\it spherically} symmetric with respect to $\psi$ and $\phi$ (similar to the azimuthal symmetry of the three-dimensional cone). We expand the Green's function as
\begin{widetext}
\begin{align} \label{Eq. expns of G in terms of g_lm}
  G_0(x,x')&= \sum_{lm} \frac{1}{\sqrt{r r'}} \frac{1}{\sqrt{\sin \theta\sin\theta'}}
  g_{lm}(r, \theta, r',\theta') \, Y_{lm}(\psi, \phi) Y^{\star}_{lm}(\psi', \phi').
\end{align}
The prefactors are introduced for later convenience. With this ansatz, Eq.~(\ref{Eq. Helmholtz eqn with 4 coordinates}) takes the form
\begin{equation} \label{Eq. for g_lm}
  -\left(\nabla^2_{(2)} -\frac{(l+1/2)^2}{r^2 \sin \theta^2}+k^2 \right) g_{lm}=\frac{\delta(r-r')}{r^2} \delta(\cos\theta-\cos\theta').
\end{equation}
Note that the Laplacian in the above equation acts in two dimensions, i.e. $\nabla^2_{(2)}=\frac{1}{r^2}\partial_r r^2 \partial_r +\frac{1}{\sin\theta}\partial_\theta\sin\theta\partial_\theta$.

Given that the boundary is of conical shape, it is convenient to break up the Green's function in the coordinate $\theta$. We should then find a completeness relation in the function space of the other variable $r$. This is obtained by the Kontrovich-Lebedev transform~\cite{Samko1993}
\begin{equation}
\nonumber
  \frac{1}{\pi}\int_{0}^{\infty} d\lambda \lambda \sinh\lambda \, k_{i\lambda-1/2}(r) k_{i\lambda-1/2}(r') =\delta(r-r'),
\end{equation}
where $k_{\nu}$ is the spherical Bessel function of order $\nu$. Using this relation, one can show that
\begin{align}\label{Eq. g0_lm in integral representation}
g_{lm}  &=\frac{\kappa}{2} (-1)^l \int_{0}^{\infty} d\lambda \lambda \, k_{i\lambda-1/2}(\kappa r) k_{i\lambda-1/2}(\kappa r')\frac{\Gamma(i \lambda+l+1)}{i \Gamma(i\lambda-l)} P_{i\lambda-1/2}^{-l-1/2}(\cos\theta_<)P_{i\lambda-1/2}^{-l-1/2}(-\cos\theta_>)\,,
\end{align}
solves Eq.~(\ref{Eq. for g_lm}).
Here, $\kappa$ is the imaginary frequency ($k=i\kappa$), $\theta_<=\min(\theta,\theta')$ and $\theta_>=\max(\theta,\theta')$. Note that this equation satisfies the Helmholtz equation in empty space. We discuss the boundary condition below. The completeness relation, together with an identity regarding the Wronskian of Legendre functions, can be exploited to see that Eq.~(\ref{Eq. g0_lm in integral representation}) indeed solves Eq.~(\ref{Eq. for g_lm}).
By a construction similar to Ref.~\cite{Maghrebi10}, we can analytically continue the complex order of the Bessel and Legendre functions to the real axis to obtain
\begin{align}
  g_{lm} & = {\kappa }  \sum_{n=l+1}^{\infty} i_{n-1/2}(\kappa r_{<}) k_{n-1/2}(\kappa r_{>})   \frac{n\,\Gamma(n+l+1)}{\Gamma(n-l)} P_{n-1/2}^{-l-1/2}(-\cos\theta) P_{-n-1/2}^{-l-1/2}(-\cos\theta').
\end{align}
Note that the asymmetry has shifted to the radial variable while the angular coordinates are treated symmetrically. We have used the symmetry $P_{n-1/2}^{-l-1/2}(-x)=(-1)^{l+n+1} P_{n-1/2}^{-l-1/2}(x)$ for integers $l< n$ to restore the latter symmetry.
Next we  take the limit $\kappa \to 0$ to find
\begin{align}\label{Eq. free Green's function in conical coordinates}
  & G_0(x,x')=\sum_{n=1}^{\infty}\sum_{l=0}^{n-1}\sum_{m=-l}^{l} \frac{ \Gamma(n+l+1)}{2\Gamma(n-l)} \frac{r_<^{n-1}}{r_>^{n+1}} \frac{P_{n-1/2}^{-l-1/2}(-\cos\theta)}{\sqrt{\sin\theta}} \frac{P_{n-1/2}^{-l-1/2}(-\cos\theta')}{\sqrt{\sin\theta'}} \, Y_{lm}(\psi, \phi) Y^{\star}_{lm}(\psi', \phi').
\end{align}
This equation provides the Green's function for the Laplace equation in empty space. For the Green's function in the presence of the conical obstacle, we modify Eq.~(\ref{Eq. g0_lm in integral representation}) to
\begin{align}\label{Eq. g_lm in integral representation}
g_{lm}  =\frac{\kappa}{2} (-1)^l  \int_{0}^{\infty} d\lambda \lambda \, k_{i\lambda-1/2}(\kappa r) k_{i\lambda-1/2}(\kappa r')&
\frac{\Gamma(i \lambda+l+1)}{i \Gamma(i\lambda-l)} \, \Big(P_{i\lambda-1/2}^{-l-1/2}(\cos\theta_<)P_{i\lambda-1/2}^{-l-1/2}(-\cos\theta_>)-\nonumber\\ & \frac{P_{i\lambda-1/2}^{-l-1/2}(\cos\Theta)}{P_{i\lambda-1/2}^{-l-1/2}(-\cos\Theta)}P_{i\lambda-1/2}^{-l-1/2}(-\cos\theta)P_{i\lambda-1/2}^{-l-1/2}(-\cos\theta')\Big)\,,
\end{align}
which vanishes when either angle is equal to $\Theta$.
It can be easily checked that this is indeed the Green's function: Upon acting with the Helmholtz operator, Eq.~(\ref{Eq. for g_lm}), the second line of the last equation gives a delta function while the third line vanishes since it has no discontinuity along $\theta=\theta'$.  Also the boundary condition is clearly satisfied as $\theta \to \Theta$, the opening angle of the cone. Next we rotate to the real axis and take the limit $\kappa \to 0$ to obtain the corresponding Green's function for the Laplace equation
\begin{align}\label{G0 cone}
G_0(x,x')=  \sum_{n=1}^{\infty} \sum_{l=0}^{\infty}\sum_{m=-l}^{l} \frac{\pi}{2} (-1)^l \frac{\Gamma(\rho_n+l+1)}{\sin(\rho_n \pi) \Gamma(\rho_n-l)} &\frac{P_{\rho_n-1/2}^{-l-1/2}(\cos\Theta)}{\partial_{\rho_n}P_{\rho_n-1/2}^{-l-1/2}(-\cos\Theta)} \frac{r_<^{\rho_n-1}}{r_>^{\rho_n+1}} \frac{P_{\rho_n-1/2}^{-l-1/2}(-\cos\theta)}{\sqrt{\sin\theta}}\times
\nonumber \\ &\frac{P_{\rho_n-1/2}^{-l-1/2}(-\cos\theta')}{\sqrt{\sin\theta'}} \, Y_{lm}(\psi, \phi) Y^{\star}_{lm}(\psi', \phi').
\end{align}
The integer $n$ represents the $n$-th root of the transcendental equation
\begin{equation}
  P_{\rho_n-1/2}^{-l-1/2}(-\cos\Theta)=0\,.
\end{equation}

The solution for $\rho_n$ also depends on $\Theta$ and $l$.
In the limit of $\Theta\to0$, i.e. for a sharp cone, it can be shown that \cite{Siegel1953}
\begin{equation}
  \rho_n \approx l+n+\frac{\Gamma(2l+n+1)}{\Gamma(l+3/2)\Gamma(l+1/2)\Gamma(n)}(\tan\Theta)^{2l+1},
\end{equation}
and the asymptotic behavior in this limit is dominated by the root for $l=0$.
To proceed further, we take advantage of the identities,
\begin{align}\label{Eq. Legendre}
  &P_{\rho-1/2}^{-1/2}(\cos\theta)=\frac{1}{\rho }\sqrt{\frac{2}{\pi {\sin\theta}}}\sin(\rho\theta),\nonumber\\
  &P_{\rho-1/2}^{-1/2}(-\cos\theta)=\frac{1}{\rho }\sqrt{\frac{2}{\pi {\sin\theta}}}\sin(\rho(\pi-\theta)),
\end{align}
the second of which dictates
\begin{equation} \label{Exp cone}
  \rho_n=\frac{n\pi}{\pi-\Theta}\,.
\end{equation}
Hence, using Eqs.~(\ref{G0 cone}-\ref{Exp cone}), we find, for a single cone,
\begin{align}
  G_0&{\large|}_{l=m=0}(x,x')= \frac{1}{4\pi^2}\sum_{n}\frac{r_<^{\rho_n-1}}{n \, r_>^{\rho_n+1}} \frac{\sin \rho_n (\pi-\theta)}{\sin\theta}\frac{\sin \rho_n (\pi-\theta')}{\sin\theta'}.
\end{align}

We can similarly find the Green's function in the space between two cones aligned along a common axis which touch at their tips.
Designating the opening angles by $\Theta_1$ and $\pi-\Theta_2$,  the available space is characterized by the polar angle $\Theta_1<\theta<\Theta_2$.
The Green's function (for the Helmholtz equation) is similar to Eq.~(\ref{Eq. g0_lm in integral representation}) with the substitution of $P_{i\lambda-1/2}^{-l-1/2}(\cos\theta_<)P_{i\lambda-1/2}^{-l-1/2}(-\cos\theta_>)$ by
\begin{align}
  &\frac{ \left(P_{i\lambda-1/2}^{-l-1/2}(\cos\theta_<)-\frac{P_{i\lambda-1/2}^{-l-1/2}(\cos\Theta_1)}{P_{i\lambda-1/2}^{-l-1/2}(-\cos\Theta_1) } P_{i\lambda-1/2}^{-l-1/2}(-\cos\theta_<)\right)
\left(P_{i\lambda-1/2}^{-l-1/2}(-\cos\theta_>)-\frac{P_{i\lambda-1/2}^{-l-1/2}(-\cos\Theta_2)}{P_{i\lambda-1/2}^{-l-1/2}(\cos\Theta_2) }P_{i\lambda-1/2}^{-l-1/2}(\cos\theta_>)\right)}{{1-\frac{P_{i\lambda-1/2}^{-l-1/2}(\cos\Theta_1)}{P_{i\lambda-1/2}^{-l-1/2}(-\cos\Theta_1)}\frac{P_{i\lambda-1/2}^{-l-1/2}(-\cos\Theta_2)}{P_{i\lambda-1/2}^{-l-1/2}(\cos\Theta_2)}}}. \nonumber \\
\end{align}

We then rotate to the real axis where we should find the roots of the transcendental equation
\begin{align}
P_{\rho-1/2}^{-l-1/2}(-\cos\Theta_1)&P_{\rho-1/2}^{-l-1/2}(\cos\Theta_2)-  {P_{\rho-1/2}^{-l-1/2}(\cos\Theta_1)}{P_{\rho-1/2}^{-l-1/2}(-\cos\Theta_2)}=0.
\end{align}
As we are mainly interested in a cone attached to a plate,  we choose $\Theta_1 \equiv \Theta$ and $\Theta_2=\pi/2$.
By rotating to the real axis and taking the limit $\kappa\to 0$, the full Green's function is then obtained as
\begin{align}
G_0(x,x')=  \sum_{n} &\sum_{l=0}^{\infty}\sum_{m=-l}^{l} \frac{\pi}{2} (-1)^l \frac{r_<^{\rho_n-1}}{r_>^{\rho_n+1}} \frac{\Gamma(\rho_n+l+1)}{\sin(\rho_n \pi) \Gamma(\rho_n-l)} \frac{P_{\rho_n-1/2}^{-l-1/2}(\cos\Theta)}{\partial_{\rho_n}\left(P_{\rho_n-1/2}^{-l-1/2}(-\cos\Theta)-P_{\rho_n-1/2}^{-l-1/2}(\cos\Theta)\right)}
\times \nonumber \\
&\frac{P_{\rho_n-1/2}^{-l-1/2}(-\cos\theta)-P_{\rho_n-1/2}^{-l-1/2}(\cos\theta)}{\sqrt{\sin\theta}}
\frac{P_{\rho_n-1/2}^{-l-1/2}(-\cos\theta')-P_{\rho_n-1/2}^{-l-1/2}(\cos\theta')}{\sqrt{\sin\theta'}} \, Y_{lm}(\psi, \phi) Y^{\star}_{lm}(\psi', \phi'),
\end{align}
where $\rho_n$ is the $n$-th root of the transcendental equation
\begin{equation}
P_{\rho_n-1/2}^{-l-1/2}(-\cos\Theta)-P_{\rho_n-1/2}^{-l-1/2}(\cos\Theta)=0.
\end{equation}
For a sharp cone, we can focus on $l=0$, in which case, using Eq.~(\ref{Eq. Legendre}), we find
\begin{equation}
  \rho_n=\frac{n \pi}{\pi/2-\Theta}\,.
\end{equation}
The Green's function for the cone+plate configuration then reads
\begin{align}
  G_0&{\large|}_{l=m=0}(x,x')= \frac{1}{4\pi^2}\sum_{n}\frac{r_<^{\rho_n-1}}{n \, r_>^{\rho_n+1}} \frac{\sin \rho_n (\pi/2-\theta)}{\sin\theta}\frac{\sin \rho_n (\pi/2-\theta')}{\sin\theta'}\,.
\end{align}
\end{widetext}

\end{document}